\documentclass[two column,prb,showpacs,multicol,amsmath,amssymb]{revtex4}
\usepackage[dvips]{graphicx}
\usepackage{graphicx}% Include figure fes
\usepackage{dcolumn}% Align table columns on decimal point
\usepackage{bm}% bold math
\usepackage{graphics}
\usepackage{epsfig,color}
\usepackage{epstopdf}
\usepackage{soul}
\begin{document}
\title {Heat dissipation and its relation to molecular orbital energies in single-molecule junctions}
\author{ Yaghoob Naimi$^{1}$ and Javad Vahedi$^{2}\footnote{Two authors have the same collaboration}$}
\address{ $^{1}$ Department of Physics, Lamerd Branch, Islamic Azad University, Lamerd, Iran.\\
$^{2}$ Department of Physics, Sari Branch, Islamic Azad University, Sari, Iran.}
\date{\today}
\begin{abstract}
We present a theoretical study of the heat dissipation in single-molecule junctions. In order to investigate the heat dissipation in the electrodes and the relationship between the transmission spectra and the electronic structures, we consider a toy model that in which electrodes linked by a two-level molecular bridge. By using of the Landauer approach, we show how heat dissipation in the electrodes of a molecular junction is related to its transmission characteristics. We show that in general heat is not equally dissipated in the left and right electrodes of the junction and it depends on the bias polarity and the positions of  molecule's energy levels with respect to the Fermi level.  Also, we exploit the C$_{60}$ molecule as a junction and the results show a good agreement with the toy model. Our results for the heat dissipation are remarkable in the sense that they can be used to detect which energy levels of a junction are dominated in the transport process.
\end{abstract}
\pacs{85.35.Ds, 85.65.+h, 81.07.Nb, 72.10.Di}

\maketitle
\section {Introduction}
In recent decades, charge transport through single-molecule and other nano-scale systems are being actively studied both theoretically and experimentally \cite{ree, red}. It is suggested that a single-molecule could be a bridge between two electrodes and behaves as circuit components \cite{mai, saf}. Theoretical perspectives of electronic transport in single-molecule junctions can play a major role in understanding and manufacturing novel nanoscale electronic devices \cite{ker, por, rei, nit}. Two essential aspects of molecular junctions are thermoelectric effects and heat dissipations that require more investigations due to experimental challenges. Thermoelectric effects involve basic interplay between the electronic and thermal properties of a system. A temperature gradient in a conductor causes charge flow and leads to a electric current. Theoretical \cite{pau,paul,ke,fin,liu,berg1,berg2,noz,que,ser,stad,saha,bil,bur,mar} and experimental \cite{red, bah, mal, tan, yee, tan2, wid, eva, wid2} works on thermopower in molecular junctions show that the thermopower reveals some useful information that they can not be obtained from the standard current-voltage measurements.  Charge carrier transport is always accompanied by heat dissipation, i.e., Joule heating. In the recent experimental work \cite{lee}, authors used custom fabricated scanning probes with integrated nanoscale thermocouples for studying heat dissipation in the electrodes of single molecular junctions. They found that when the transmission through the junction strongly depends on energy, the heat dissipation in electrodes is unequal and shows asymmetric behaviors. Furthermore, they showed that the heat dissipation depends on both bias polarity and the identity of the majority charge carriers. Also, recently in the theoretical work \cite{zot} authors expressed the basic principle that govern the heat dissipation in molecular junctions in more detail. They showed how the heating in electrodes of a molecular junctions is determined by its electronic structure. They concluded that heat is asymmetrically dissipated in electrodes of molecular junctions and it depends on the bias polarity. Some of their results obtained by analyzing the heat dissipation in single-level molecular junctions.

Motivated by ref. \cite{zot}, we report here a detailed theoretical analysis of the joule heating in current-carrying single-molecule junctions. In spite of previous mentioned work, we consider a two-level molecule (toy model) as a bridge between two metal electrodes, so each electrode couples to a molecule with two branches as Fig. 1 shows it. We show how the heating in the electrodes of a molecular junction is determined by its electronic structure. To this end, we first obtain exact expression for a transmission function of a toy model by using of the generalized Green's function formalism and then by using of transmission we obtain the relations of heat dissipation in both electrodes. We verify that in these structures, the energies of the molecular orbitals, in particular, the highest occupied
molecular orbital (HOMO) and the lowest unoccupied molecular orbital (LUMO), have a vital role for the
electronic transport through single organic molecules. We show that the three selections for two energy levels as LUMO-LUMO, LUMO-HOMO and HOMO-HOMO lead to different heat dissipations in electrodes. Furthermore, we show that in contrast with the single-level model used in \cite{zot} for benzene-based molecules, heat dissipation in our model is not always asymmetric. In the next step, as another molecular bridge, we consider the $C_{60}$ molecule that sandwiched between two metal electrodes (Au) via single and multiple contacts. Our numerical conclusions for $C_{60}$, by using of effective single-particle tight-binding model, is well matched with results of the toy model.

The rest of the paper is organized as follows. In Sec. II we define a toy model and present a detailed  formalism  for finding  its transmission and its heat dissipation characteristics. In Sec. III we present our numerical results for the toy model. Sec. IV is assigned for numerical results  of a real Au-$C_{60}$-Au molecular junction based on tight-binding model. In Sec. V we summarize our findings and some technical details are presented in the Appendices.
%########################################################################
%######################                        ##########################
%######################      Section II       ##########################
%######################                        ##########################
%#######################################################################
\section {Theory and model}
We consider the molecule (introduced as a set of energy levels) placed in between two electrodes (left $L$ and right $R$) and plays a role of channel. The electrodes are behaved as free electron reservoirs with approximately continuous energy spectra. The electronic transport properties of molecular junctions are
govern by quantum mechanical laws. One of most important framework for studying  theoretical nanoelectronics is a Landauer frameworks \cite{butt}. Landauer approach is based on the description of electron transport through elastic scattering model. The thermalized electrons in reservoirs will be scattered when they come into the
channel, but their transport are completely coherent between the electrodes. One can interpret the conductance of channel as an elastic scatterer, by the quantum mechanical probabilities of transmission $T(E)$, that corresponds to electrons with energy $E$.

\subsection{Transmission and heat dissipation in two-level (toy) model}
%#########################################################
%######################                        ##########################
%######################     Figure1            ########################
%######################                        ##########################
%#########################################################

\begin{figure}
 \centering
 \includegraphics[width=.9\columnwidth]{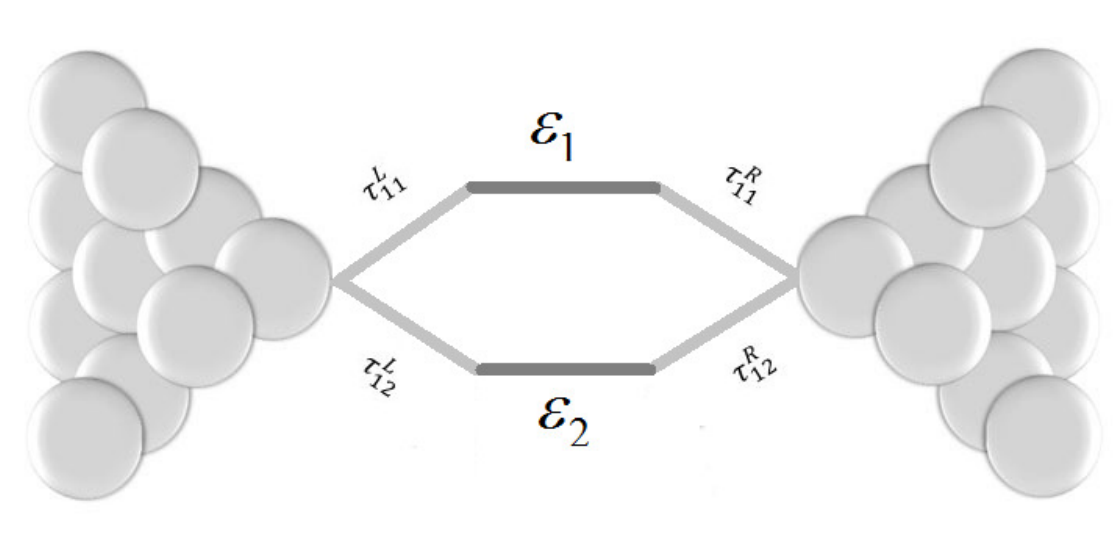}
 \caption{Schematic representation of the two-level model.}
\label{fig1}
\end{figure}
%#########################################################
%######################                        ##########################
%######################     Figure2            ########################
%######################                        ##########################
%#########################################################
\begin{figure}
  \includegraphics[width=0.8\columnwidth]{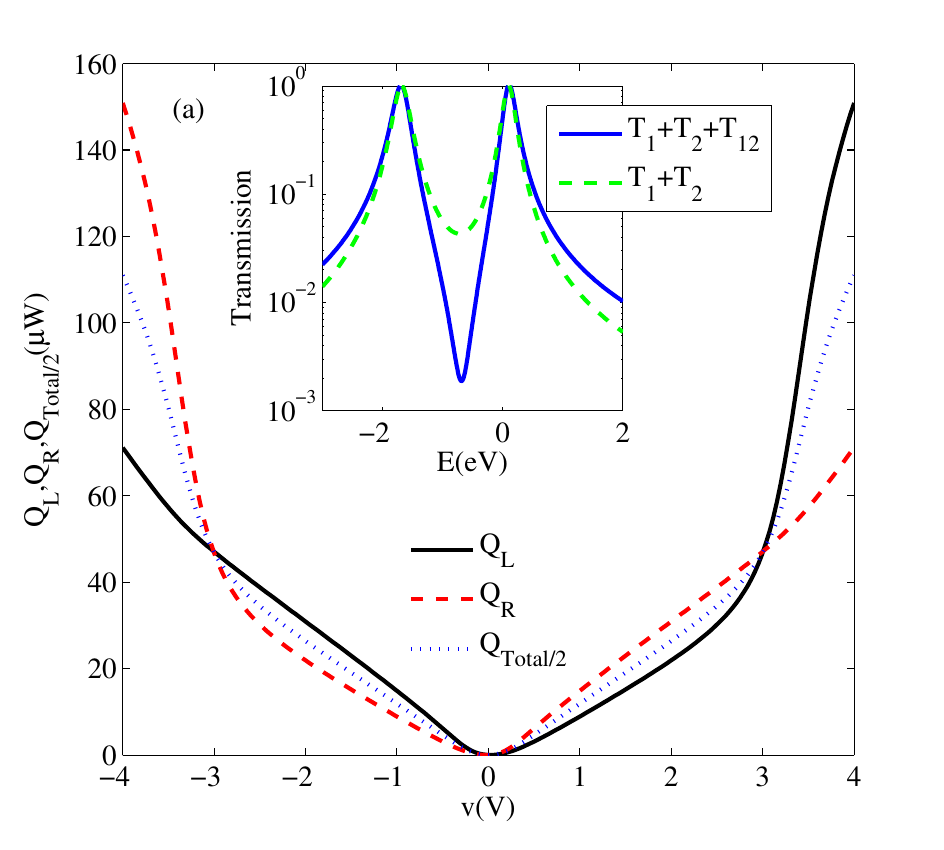}
  \includegraphics[width=0.8\columnwidth]{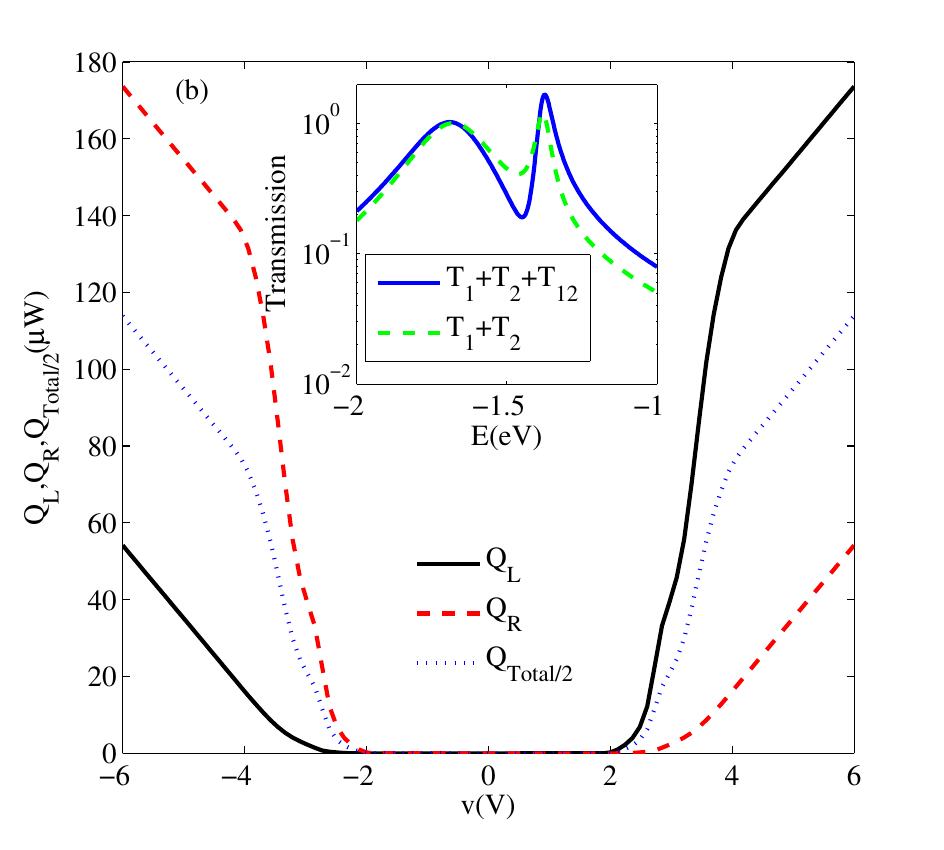}
  \includegraphics[width=0.8\columnwidth]{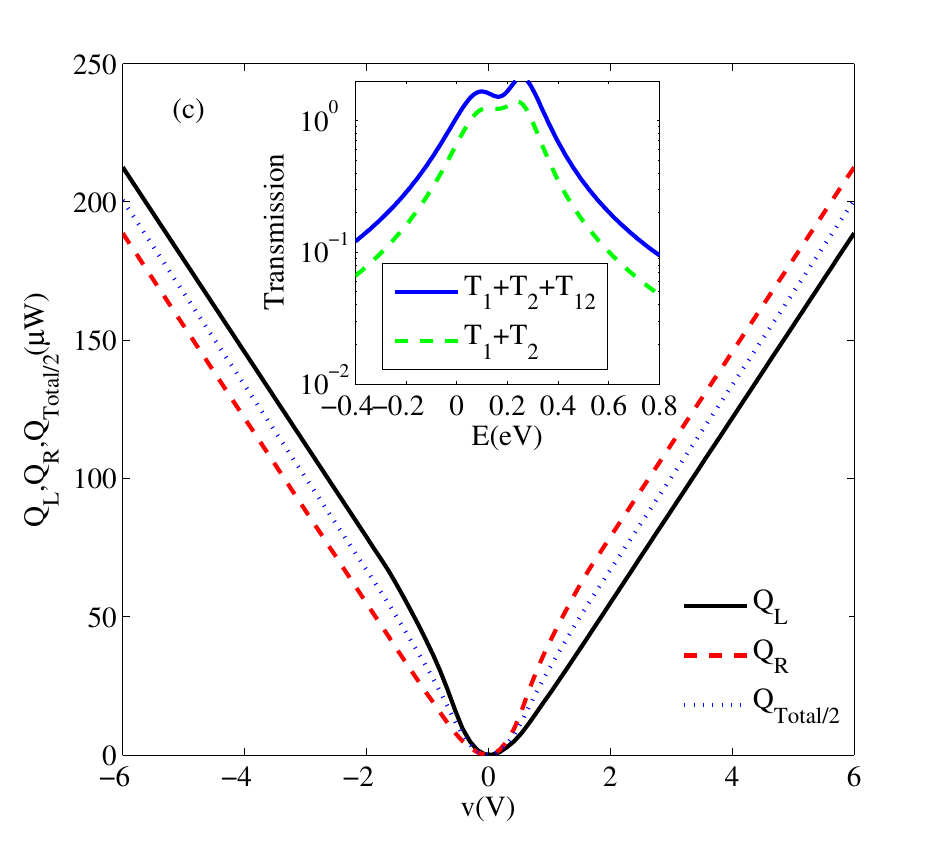}
 \caption{(Color online) The power dissipated in the left electrode $Q_L$  (black or solid line),  the right electrode $Q_R$ (red or dashed line) and the half of total power dissipated  $Q_{Total}/2$ (blue or solid line) as a function of bias voltage. (a) upper panel for case HL, (b) middel panel for case HH and (c) bottom panel for case LL.}
\label{fig2}
\end{figure}
%#########################################################
%######################                        ##########################
%######################     Figure3          #######################
%######################                        ##########################
%#########################################################
\begin{figure}[h!]
 \includegraphics[width=1\columnwidth]{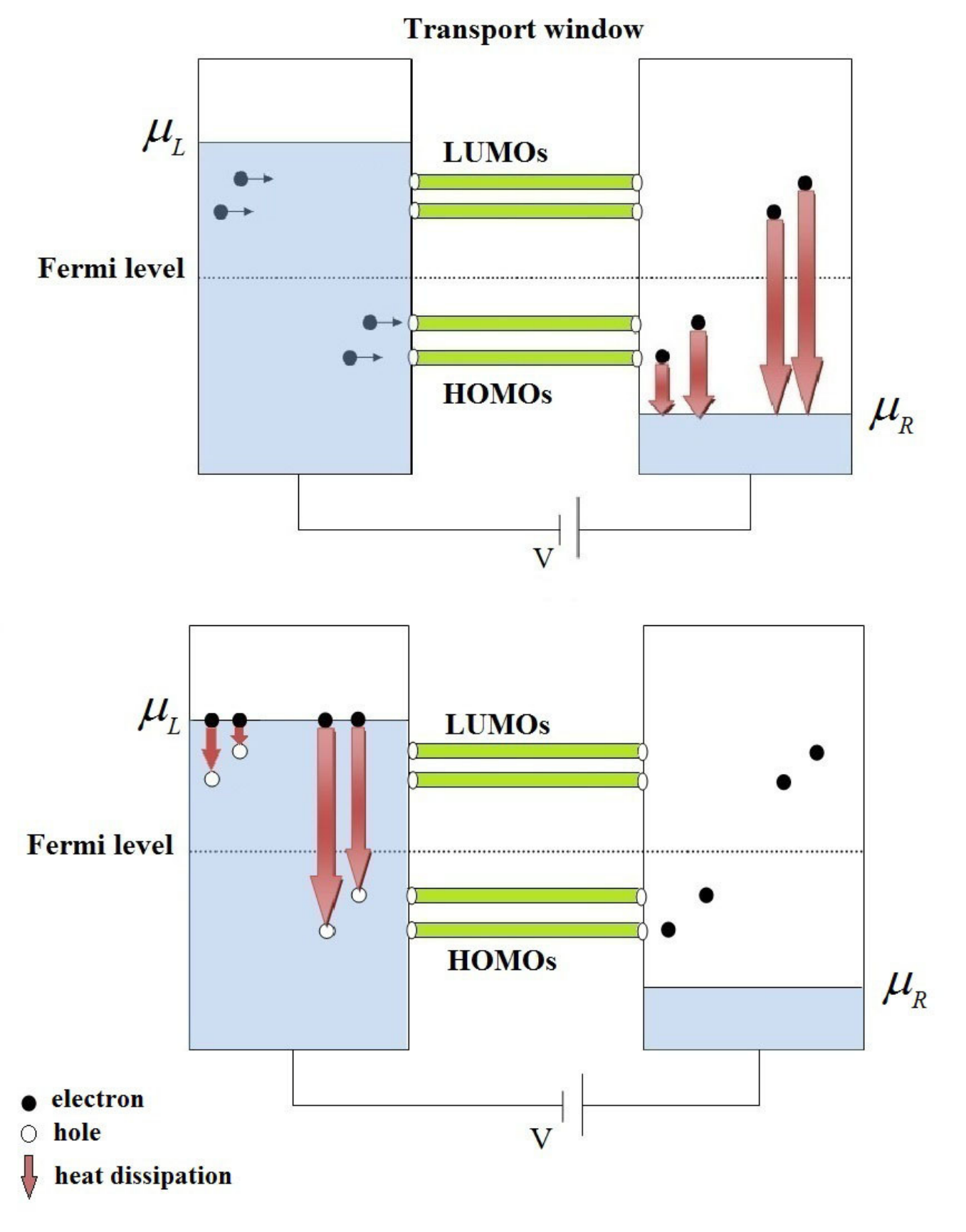}
 \caption{(Color online) (a) The $Q_L-Q_{Total}/2$ as a function of the bias. It is indicated that for bias about $|V_0|$=3V, $ Q_L=Q_{Total}/2$, i.e., symmetric heat dissipated in both electrodes and for $|V_0|>3V$ the left electrode has a major portion in total heat dissipation. (b) The power dissipated in the left lead as a function of the total power for positive and negative bias. The dashed line correspond to the power dissipated in a symmetric situation, i.e. $Q_L(V ) = Q_{Total}/2$. There is an intersection point that in which the heat dissipated in left lead is equal for positive and negative bias and after this point for positive (negative) bias the left electrode has a major (minor) portion in total heat dissipation.}
\label{fig3}
\end{figure}
%%%%%%%%%%%%%%%%%%%%%%%%%%%%%%%%%%%%%%%%%%%

Let us consider a two-terminal system with $L$ and $R$ electrodes linked
by two-level molecular junction model as Fig.\ref{fig1}.
For this model, the transmission function has been calculated in \cite{gera} and authors showed that the transmission can be expressed as sum of three terms as following
\begin{equation}\label{12}
 T(E)=T_1(E)+T_2(E)+T_{12}(E),
\end{equation}
where for a system with energies $\varepsilon_i$s and level broadening $\gamma_i$s, $T_i$s have a Lorentzian definition
\begin{equation}\label{13}
 T_i(E)=\frac{\gamma_i^2}{(E-\varepsilon_i)^2+\gamma_i^2},\ \ \ \ \ \ \ i=1, 2
 \end{equation}
 and interference enters via
 \begin{equation}\label{13}
T_{12}(E)=2\gamma_1\gamma_2\frac{(E-\varepsilon_1)(E-\varepsilon_2)+\gamma_1\gamma_2}{\left((E-\varepsilon_1)^2+\gamma_1^2\right)\left((E-\varepsilon_2)^2+\gamma_2^2\right)} \end{equation}
We can extend the $T_{12}$  according to $T_{i}$s as
   \begin{equation}\label{25}
T(E)=(1+ C_1E+ C_2)T_1(E)+(1+ D_1E+ D_2)T_2(E),
   \end{equation}
   where constants $C_i$s have a following definitions
   \begin{eqnarray}
     C_1 &=& \frac{2\gamma_2(\varepsilon_1-\varepsilon_2)}{\gamma_1[(\varepsilon_1-\varepsilon_2))^2+(\gamma_1+\gamma_2)^2]} \\
     C_2 &=& 2\gamma_2\frac{\varepsilon_1\varepsilon_2+\gamma_1\gamma_2+\gamma_1^2-\varepsilon_1^2}{\gamma_1[(\varepsilon_1-\varepsilon_2))^2+(\gamma_1+\gamma_2)^2]}
       \end{eqnarray}
   and constants $D_i$s will obtain by applying  $\varepsilon_1\leftrightarrow\varepsilon_2$ and also $\gamma_1\leftrightarrow\gamma_2$ exchanges in corresponding $C_i$s.
One, who is interested in the details of the above calculation, can refer the appendices $A$ and $B$. 
\par
Let us suppose that a voltage bias $V$ is applied across the system. It causes that the electrochemical potential of the left and right electrodes shifts such that $\mu_L-\mu_R = \pm eV$, where $e>0$ is the electron charge. We call positive (negative) bias for plus (minus) sign. On the other words, for positive (negative) bias electrons flow from the left (right) electrode to the right (left) electrode.  The fact that elastic scattering is not associated with any energy loss in the junction regions implicitly reminds that when an electron of energy $E$ tunnels from the left to the right (in positive bias) releases its excess energy $E-\mu_R$ in the right lead, while the hole, that is created behind the electron, is filled up releasing an energy equal to $\mu_L-E$ in the left lead. More precisely, these released energies dissipate as a heat in two leads. According to the calculations have been carried out in the appendix $C$ or in \cite{zot}, the rate of heat dissipation (heat per unit of time) in left and right electrodes and the total heat dissipation are
\begin{eqnarray}
 Q_L(V)&=&\frac{2}{h}\int_{-eV/2}^{+eV/2}(eV/2-E)T(E)dE,\\
 Q_R(V)&=&\frac{2}{h}\int_{-eV/2}^{+eV/2}(E+eV/2)T(E)dE, \\
 Q_{Total}(V)&=&\frac{2eV}{h}\int_{-eV/2}^{+eV/2}T(E)dE,
\end{eqnarray}
By inserting transmission function (1) in Eq.(9) and taking the integral,
we obtain the explicit expression for the total power dissipation in two-level model as
\begin{widetext}
\begin{eqnarray}
  \nonumber Q_{Total}(V) &=& \frac{2eV\gamma_1}{h}(1+\varepsilon_1C_1+C_2)\left[\arctan(\frac{eV/2-\varepsilon_1}{\gamma_1})+\arctan(\frac{eV/2+\varepsilon_1}{\gamma_1})\right] + \frac{eV\gamma_1}{h}C_1\ln\left[\frac{\gamma_1^2+(eV/2-\varepsilon_1)^2}{\gamma_1^2+(eV/2+\varepsilon_1)^2}\right] \\
 &+& \mbox{first and second terms with} \left\{C\leftrightarrow D, \varepsilon_1\leftrightarrow \varepsilon_2, \gamma_1\leftrightarrow\gamma_2\right\}
\end{eqnarray}
\end{widetext}
%#########################################################
%#########################################################
%######################                        ##########################
%#########################################################
\section{Numerical results for two-level model}
 Here, we present our numerical  results for the power dissipated in the two-level molecular junction with  $Au$-electrodes  as a function of applied bias. The two levels that one should refer to here is derived from the C$_{60}$-HOMO quintett and LUMO-triplet \cite{dif}, that we will obtain them in the next section when we consider C$_{60}$ as a molecular junction. Accordingly, we define three different cases based on which levels contribute  during the transmission:
\begin{itemize}
  \item Case HL: HOMO-LUMO, one of the levels is belong to HOMO and the other is belong to LUMO.
  \item Case HH: HOMO-HOMO, both levels are belong to HOMOs.
  \item Case LL: LUMO-LUMO, both levels are belong to LUMOs
\end{itemize}
Although, we know that in \cite{kim}, the measurements of the Seebeck coefficients were claimed to indicate that, in Au-C$_{60}$-Au junctions, the transport is dominated by the LUMO and in \cite{lu} it was shown, via STM spectroscopy,  that when C$_{60}$ was adsorbed on a flat Au surface the Fermi level was clearly closer to the LUMO than to the HOMO; but we consider three above cases to show that heat dissipation can help us to trace the energy levels that contribute during transmission.
Values of the energy level $\varepsilon_i$, and broadening $\gamma_i$ are given in Table I.
\begin{table}[ht]
\caption{Energy levels, $\varepsilon_i$ and levels broadening, $\gamma_i$, for C$_{60}$ } % title of Table
\centering % used for centering table
\begin{tabular}{c c c c c} % centered columns (4 columns)
\hline\hline \\%inserts double horizontal lines
$E_f=0$\\
\hline\\
Case&$(\varepsilon_1, \gamma_1$)eV&$(\varepsilon_2, \gamma_2)$eV\\
\hline % inserts single horizontal line
HL &(-1.68, 0.149)&(0.1, 0.116)\\ % inserting body of the table
HH &  (-1.68, 0.149) & (-0.138, 0.026) \\
LL &  (0.1, 0.116)& (0.25, 0.08)\\
 [1ex] % [1ex] adds vertical space
\hline %inserts single line
\end{tabular}
\label{table:nonlin} %is used to refer this table in the text%
\end{table}
The values in table I have been computed in \cite{gera} by using DFT. In all three cases, the power dissipation in left and right electrodes is asymmetric and the condition (C9) is verified. A detailed elaboration of Fig.\ref{fig2} (a) shows a cross point between the power dissipated in left and right electrodes, $Q_L(\pm V)=Q_R(\pm V) $, which it does not happen in the other two models. This cross point means that heat is equally dissipated in both electrodes. To further investigate of the portion of the left electrode in heat dissipation in case HL, we plotted the $ Q_L-Q_{Total}/2$ as a function of the bias in Fig.\ref{fig3}.  Panel (a) of this figure shows that, there is a given positive bias about $V_{0}=3 V$ in which $ Q_L-Q_{Total}/2=0$ or $ Q_L=Q_{Total}/2$, i.e., symmetric heat dissipated in both electrodes. More precise, for the  positive bias when $0<V<V_0$ ($V>V_0$ ) we find $Q_L<Q_{Total}/2$ ($Q_L>Q_{Total}/2$), so the portion of the left electrode in heat dissipation is less (more) than half of the total heat. This behavior is exactly reversed for negative bias. Briefly, the main conclusion of these results reveals when we plot $Q_L$ as a function of $Q_{Total}$ for both positive and negative biases in panel (b). There is a point of intersection that in this point the heat dissipation in the left lead is equal for positive and negative bias and after this point for positive (negative) bias the left electrode has a major (minor) portion in the total heat dissipation, although before the intersection point, there is no significant difference between the heat dissipation for the positive and negative bias.
%#########################################################
%######################                       ##########################
%######################     Figure4           ########################
%######################                       ##########################
%#########################################################
%%%%%%%%%%%%%%%%%%%%%%%%%%%%%
\begin{figure}
 \includegraphics[width=1.1\columnwidth]{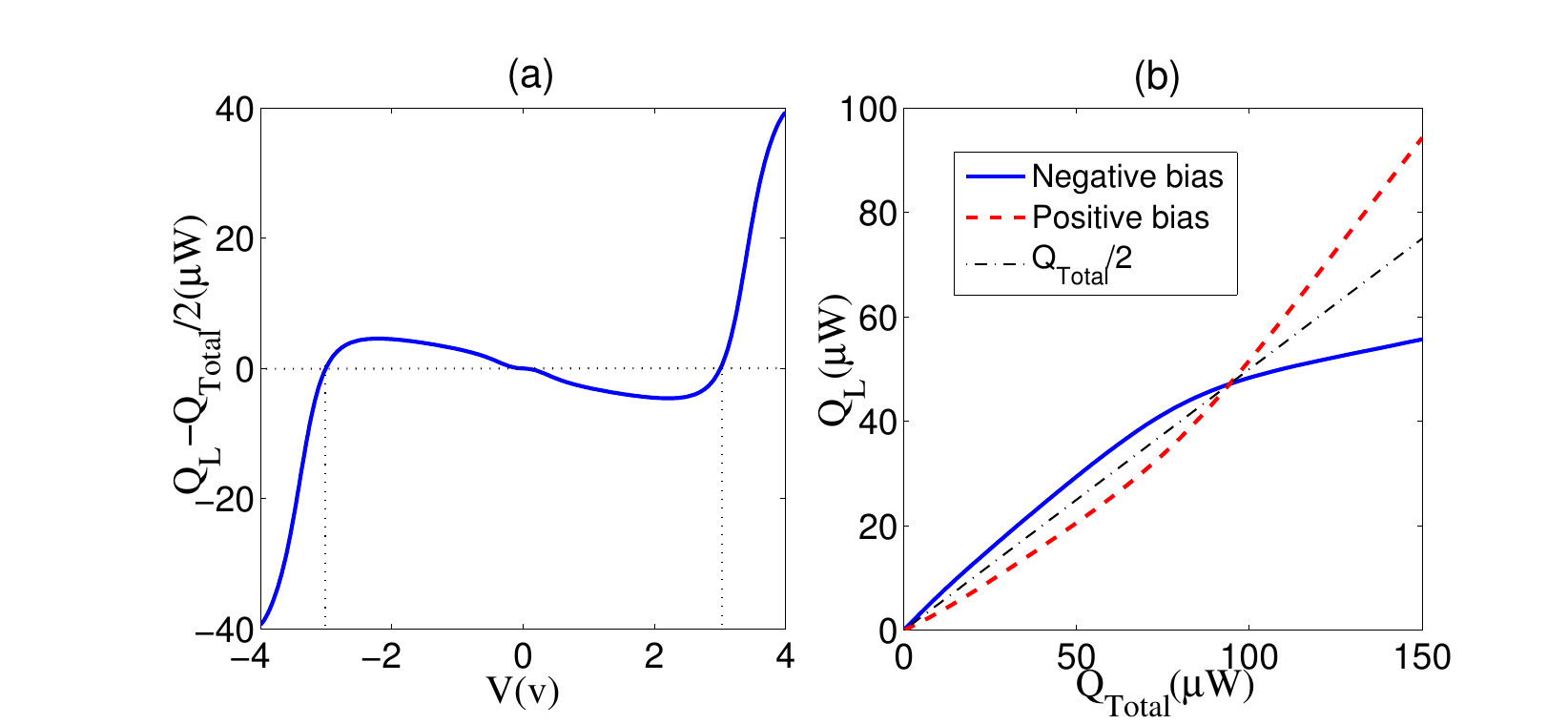}
 \caption{(Color online)  Schematic representation of the power dissipated in a molecular junction.
 After applying bias voltage transport window opens and molecular orbitals in the window can participate
 on the transmission processes. Now an electron tunnels from the left electrode to the right electrode,  so
 releases its excess energy $E-\mu_R$ in the right electrode.  For those electrons transmitted via the LUMOs more power is dissipated in the right electrode (see the upper panel). Transmitted electrons leave holes in the left electrode, so the hole is filled up by dissipating an energy equal to $\mu_L-E$ in the left electrode.  For the hole left behind by electron  transmitted via the HOMOs  more power is dissipated in the left electrode (see the bottom panel).}
\label{fig4}
\end{figure}
%%%%%%%%%%%%%%%%%%%%%%%%%%%%%%%%%%%%%%%%%%

If we focus on cases HH and LL, their behaviors can be related to the HOMO and LUMO levels domination during the electron transmission processes.  Indeed, after turning bias voltage the electrochemical potentials of the left and the right electrodes are shifted and  an energy window opens for electrons to cross the junction and it results in a net electron current in the junction. When the transport window is open through LUMOs, more (less) power dissipates in right (left) electrode (case HH and see Fig.\ref{fig2} (b) and when the transport window allows electrons to cross along the HOMOs less (more) power dissipates in left (right) electrode (case LL and see Fig.\ref{fig2} (c).  A cartoon to present the scenario is depicted in Fig.\ref{fig4}.
\par
 %#########################################################
%######################                        ##########################
%######################     Figure5            ########################
%######################                        ##########################
%#########################################################
\begin{figure}
\includegraphics[width=1.1\columnwidth]{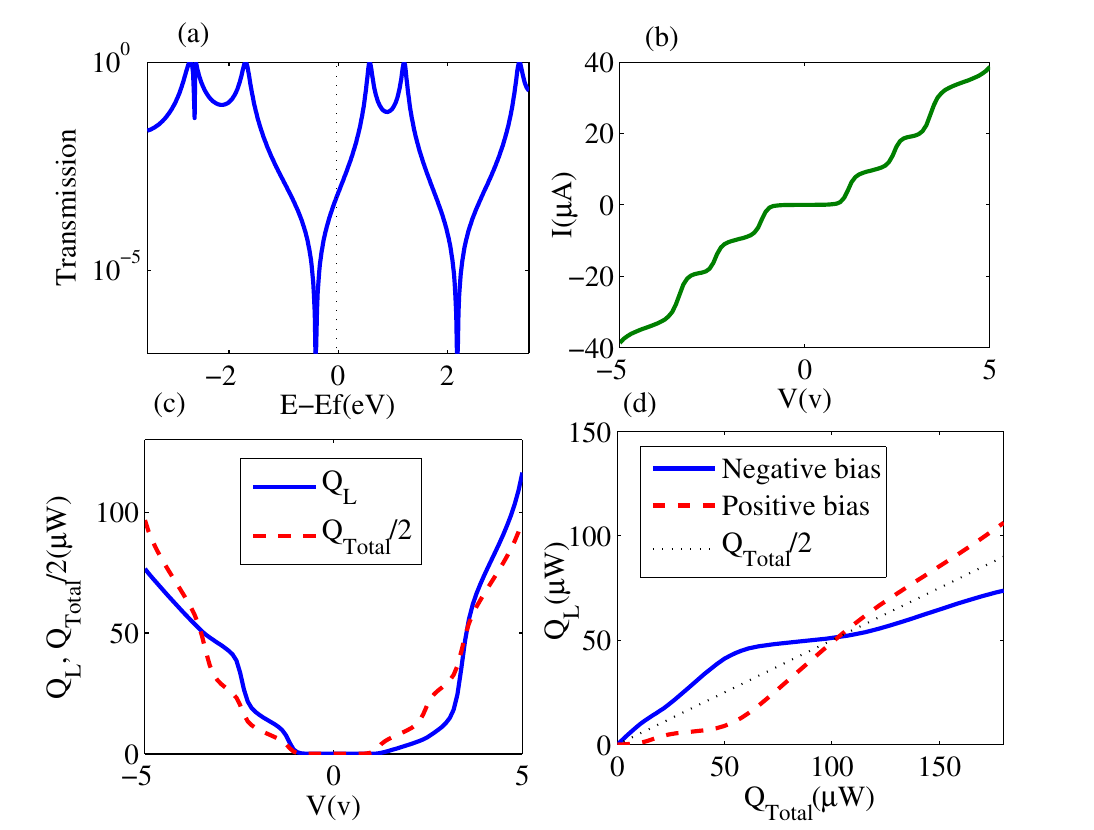}
\caption{(Color online) (a) Transmission probability as a function of energy for one-contact. (b) Current vs voltage characteristics. (c) Half of the total power dissipated in the junction and the power dissipated in the left lead as a function of the bias. (d) The power dissipated in the left lead as a function of the total power for positive (dashed line) and negative (solid line) bias. The dotted line correspond to the power dissipated in a symmetric situation, i.e. $Q_L(V)=Q_{Total}/2$.}
\label{fig5}
\end{figure}
%#########################################################
%######################                        ##########################
%######################     Figure6            ########################
%######################                        ##########################
%#########################################################
 \begin{figure}
\includegraphics[width=1.1\columnwidth]{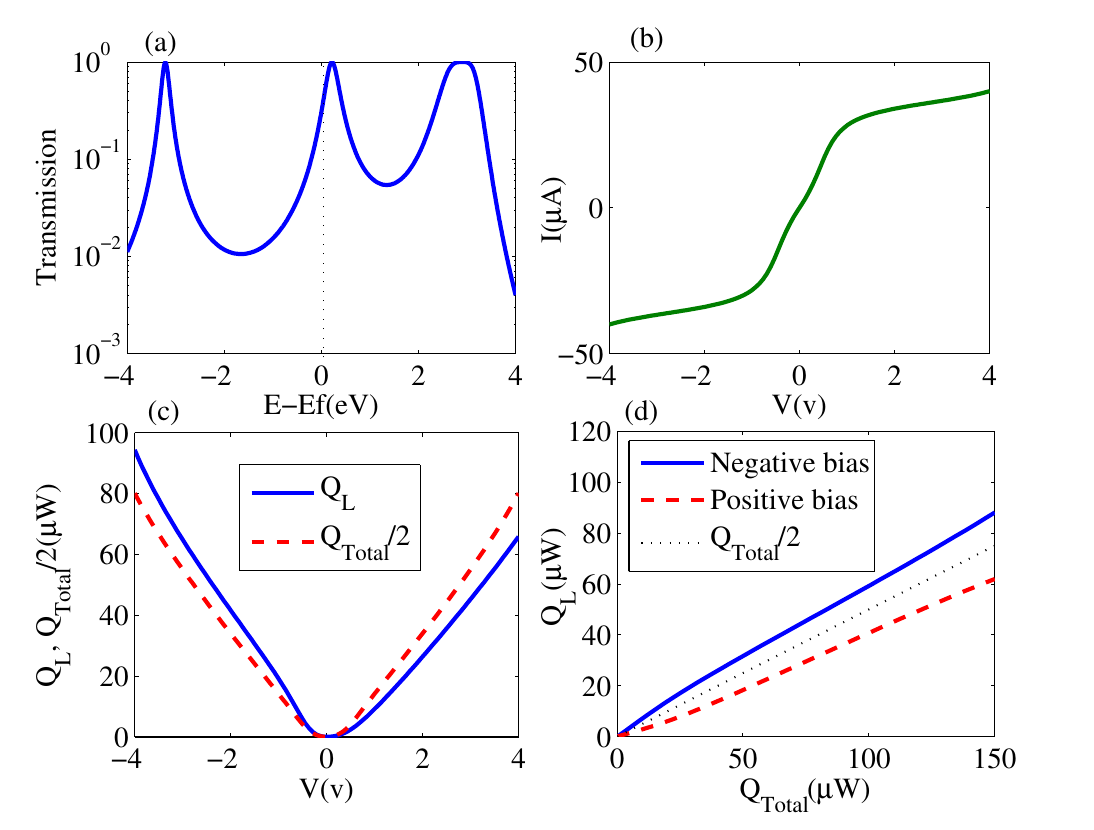}
\caption{(Color online) (a) The same description as Fig. 5 but for five-contact.}
\label{fig6}
\end{figure}
%#########################################################
%######################                        ##########################
%######################      Section III       ########################
%######################                        ##########################
%#########################################################
\section{Heat dissipation in the $C_{60}$ molecule junctions}
 We consider a system consists of a $C_{60}$ molecule attached to one-dimensional gold ($Au$) electrodes. The whole system is described within a single electron picture by a tight-binding Hamiltonian with nearest-neighbor hopping approximation. The Hamiltonian representing the entire system can be written as $\emph{H}=H_L+V_L+H_M+H_R+V_R $  where $H_{L/R}$ and $H_M$ represent the left/right electrode and the $C_{60}$  molecule, respectively.  $V_{L/R}$ defines the coupling between left/right electrodes and the $C_{60}$ molecule. In our numerical calculation we used the well-konw Newns-Anderson model for gold ($Au$) electrodes, whose self-energies are given by $\Gamma_{L/R}=V_{L/R}g_{L/R}(E)V_{L/R}^\dagger$, where $g_{L/R}(E)$ is the surface Green's function defined as  $g_{L/R}(E)=i\exp{ika}/t_{L/R}$\cite{saf, noz2}.  $t_{L/R}$ is the nearest-neighboring hopping integral in the left/right electrodes.  The hopping strength in $C_{60}$ molecule depends on the dimerization of the carbon-carbon bonds, thus we consider different hopping integral elements: $t_1$ for the single bonds and $t_2$ for the double bonds.  In the numerical calculations we set $ t_1= 2.5 eV\cite{saf},  t_2= 1.1 t_1, t_{L/R}=t_1,  E_F= 0.0 eV, $ and $T = 300 K$.

Results for one and five-contact have depicted in Fig.(\ref{fig5}) and Fig.(\ref{fig6}), respectively. In Fig.\ref{fig5} (a) we have plotted the logarithmic scale of transmission function versus the energy of a molecular junction with one contact. For an electron with energy $E$ that comes from the left connection, the probability  of transmission function reaches its saturated value (resonance peaks) for the specific energy values.
 In order to verify the heat dissipation feature in the junction, in Fig.\ref{fig5} (d) we present the power dissipated in the left lead as a function of the total power dissipated for negative and positive biases. Our results are in good agreement with case HH (Fig. \ref{fig3} (b)) of the simple two-level model. It can be seen that an intersection occurs between the power dissipated in the left lead for the negative and positive bias. Here, one can conclude that the HUMO-LUMO  have essential roles in the electron transport. In Fig.\ref{fig6} we have considered $C_ {60} $ molecule connected to the leads via its pentagon face. The result shows that the transmission function does not display any crossing point between the power dissipated in the left and right electrodes. Therefore, in this case, the power dissipated in the electrodes shows the asymmetric feature in the whole range of applied bias.  By comparing panels (c) of Fig. \ref{fig6}  and Fig. \ref{fig2}, we see that they are similar and so we conclude that the LUMOs have major roles in the electron transport. %#############################################################
%######################                                 ##########################
%######################      CONCLUSION       #########################
%######################                                 ##########################
%############################################################
\section{Conclusion}
 In summary, we have studied the heat dissipation in single-molecule junctions. Using the generalized Green's function technique and the Landauer formalism, we have presented a detailed theoretical and numerical analysis of  heat dissipation in the simple two-level toy model. We have shown how the transmission  characteristics can affect on heat dissipation in the two-terminal molecular junction. We verified that the energies of the molecular orbitals, in particular, HOMO and LUMO, play a essential role in the electronic transport through molecular junctions. We showed that the different selections of HOMO and LUMO for two energy levels lead to different heat dissipations in electrodes. Also, we have simulated C$_{60}$ molecule sandwiched between Au electrodes based on tight-binding model and Landauer approach. In order to have a different transmission spectrum for an equal transport window we have considered both single and pentagon contact cases.  We reached in good agreement between the behaviors of toy model and C$_{60}$ molecular junctions.  Indeed, by comparing numerical results of the toy model and the C$_{60}$ molecular junctions, one can find out which energy levels are dominated in the transport process.
%############################################################
%######################                        #############################
%######################      ACKNOWLEDGMENTS   ######################
%######################                        #############################
%############################################################
\section{Acknowledgements}
We are very grateful to unknown referee for his valuable comments and suggestions.
%#############################################################
%######################                            ############################
%######################      Appendix        ############################
%######################                           ############################
%#############################################################
\appendix
\section{Transmission for two-level model} \label{App:AppendixA}
The $(2\times2)$ matrix for an inverse Green function defines as
\begin{equation}\label{1}
G_{l,k}^{-1}=(E-\epsilon_{l})\delta_{l,k}-\Sigma^{L}_{l,k}(E)-\Sigma^{R}_{l,k}(E)
\end{equation}
where the elements of $(2\times2)$ self-energy matrices define as follows
 \begin{equation}\label{2}
   \Sigma_{l,k}^{\alpha}(E)=\sum_{m}\tau_{m,l}^{\alpha,*}g^{\alpha}(E)\tau^{\alpha}_{m,k},\ \ \ \ \ \ \alpha=L, R.
 \end{equation}
in which $\tau^{\alpha}_{l,m}$ denotes the hybridization matrix elements
that connect the two-level, up and down, with the electrodes; $g^{\alpha}(E)$ is a complex valued
surface Green's function of the uncoupled
leads, i.e., the left and right semi-infinite leads. The coupling matrix $\Gamma (E)$ also known as the broadening function, is related to the self-energies through
\begin{equation}\label{3}
  \Gamma^{\alpha}_{l,k} (E)=i(\Sigma^{\alpha}_{l,k}-\Sigma^{\alpha,*}_{l,k}),
\end{equation}
that from Eq. (A2), one can find
\begin{equation}\label{4}
  \Sigma^{\alpha,*}_{l,k}=\frac{g^{\alpha,*}(E)}{g^{\alpha}(E)}\Sigma^{\alpha}_{l,k}
\end{equation}
so the Eq. (A3) can be written as
\begin{eqnarray}\label{5}
 \nonumber \Gamma^{\alpha}_{l,k} (E) &=& i\left(1-\frac{g^{\alpha,*}(E)}{g^{\alpha}(E)}\right)\Sigma^{\alpha}_{l,k} \\
  &=& 2(\Im g^{\alpha}(E))\sum_{m}\tau_{m,l}^{\alpha,*}\tau^{\alpha}_{m,k}
\end{eqnarray}
the imaginary part of surface Green's functions is related to the contact density of states as
\begin{equation}\label{6}
  \varrho^{\alpha}(E)=-\frac{1}{\pi}\Im g^{\alpha}(E)
\end{equation}
so we obtain
\begin{equation}\label{7}
  \Gamma^{\alpha}_{l,k} (E)=2\pi \varrho^{\alpha}(E)\sum_{m}\tau_{m,l}^{\alpha,*}\tau^{\alpha}_{m,k}
\end{equation}
One can calculate the transmission from the Green's function method, using the relation
\begin{equation}\label{8}
  T(E)=Tr\{\Gamma^{L}G\Gamma^{R}G^{\dag}\}
\end{equation}
by using of relation (A7), the above equation converts to
%%%%%%%%%%%%%%%%%%%%%%%%%%%%%%%%%%%%%%%%%%%
%%%%%%%%%%%%%%%%%%%%%%%%%%%%%%%%%%%%%%%%%%%
%%%%%%%%%%%%%%%%%%%%%%%%%%%%%%%%%%%%%%%%%%%
\begin{equation}\label{9}
  T(E)=4\pi^2 \varrho^{L}\varrho^{R}Tr(\tau^{L,\dag}\tau^{L} G\tau^{R,\dag}\tau^{R} G^{\dag})
\end{equation}
since the trace is invariant under cyclic permutations, i.e., $Tr(ABCDEF)=Tr(BCDEFA)$, so we can rewrite the above equation as
\begin{equation}\label{10}
   T(E)=4\pi^2 \varrho^{L}\varrho^{R}Tr(\tau^L G \tau^{R,\dag}\tau^{R} G^{\dag}\tau^{L,\dag})
\end{equation}
because $\tau^{\alpha}=(\tau^{\alpha}_{11},\tau^{\alpha}_{12})$ is $(1\times2)$ matrix, so the multiplied $\tau^{L} G \tau^{R,\dag}\tau^{R} G^{\dag}\tau^{L,\dag}$ is a $(1\times1)$ matrix and equation (10) simplifies as
\begin{eqnarray}\label{11}
\nonumber T(E) &=& 2\pi^2 \varrho^{L}\varrho^{R}(\tau^{L} G \tau^{R,\dag})(\tau^{R} G^{\dag}\tau^{L,\dag})^{\dag} \\
&=& 2\pi^2 \varrho^{L}\varrho^{R}|\tau^{L} G \tau^{R,\dag}|^2,
\end{eqnarray}

Suppose a representation with unitary matrix $U$ which diagonalizes
the Green function $G$ to $G^{d}$ as
\begin{equation}\label{14}
  G^{d}=U^{-1}GU=\left(
                   \begin{array}{cc}
                     \frac{1}{E-z_1} & 0 \\
                     0& \frac{1}{E-z_2} \\
                   \end{array}
                 \right)
\end{equation}
Thus equation (A11) can be written as
\begin{equation}\label{15}
   T(E)=|\eta^{L}G^{d}\eta^{R,\dag}|^2
\end{equation}
This representation will not necessarily diagonalize $\Gamma^{\alpha}$, indeed, as we will show, quantum interference
effects often arise from the non-diagonal elements $\Gamma^{\alpha}$ . In this representation,
the  effective hybridization matrices are
\begin{equation}\label{16}
 \eta^{L}=\sqrt{2\pi\varrho^{L}}\tau^{L} U\ \ \ \ \ \eta^{R,\dag}=\sqrt{2\pi\varrho^{R}}U^{-1}\tau^{R,\dag}
\end{equation}
and $\Gamma^{\alpha}$ is non-diagonal as
\begin{equation}\label{17}
 U^{-1}\Gamma^{\alpha}U=\eta^{\alpha,\dag}\eta^{\alpha}=\left(
                                                          \begin{array}{cc}
                                                            |\eta^{\alpha}_{11}|^2 & \eta^{\alpha,*}_{11}\eta^{\alpha}_{12} \\
                                                            \eta^{\alpha}_{11}\eta^{\alpha,*}_{12} & |\eta^{\alpha}_{12}|^2 \\
                                                          \end{array}
                                                        \right)
 ,
\end{equation}
The expansion of transmission according to effective hybridization matrices elements is
 \begin{equation}\label{16}
   T(E)=\left|\frac{\eta^{L}_{11}\eta^{R,*}_{11}}{E-z_1}+\frac{\eta^{L}_{12}\eta^{R,*}_{12}}{E-z_2}\right|^2
 \end{equation}
the above equation can be expressed with three following terms
\begin{equation}\label{17}
  T(E)=T_1(E)+T_2(E)+T_{12}(E),
\end{equation}
the two first terms constitute the non-mixing contributions
from each energy level
\begin{equation}\label{18}
  T_1(E)=\frac{|\eta^{L}_{11}|^2|\eta^{R}_{11}|^2}{|E-z_1|^2},\ \ \ \ T_2(E)=\frac{|\eta^{L}_{12}|^2|\eta^{R}_{12}|^2}{|E-z_2|^2}
\end{equation}
Interference enters via the third term as
\begin{equation}\label{19}
  T_{12}(E)=2\Re\left(\frac{\eta^{L}_{11}\eta^{L,*}_{12}\eta^{R,*}_{11}\eta^{R}_{12}}{(E-z_1)(E-z_2)^*}\right)
\end{equation}
\section{Level broadening}
According to Eq. (A12) it is straightforward that we have
\begin{eqnarray}
  \nonumber det(G^{-1})&=&\frac{1}{det(G^d)}=\frac{1}{det(G)}\\
  &=&(E-z_1)(E-z_2)      \\
  \nonumber Tr(G^{-1})&=&\frac{1}{det(G)}Tr(G)=\frac{1}{det(G)}Tr(G^d)\\
  &=&2E-(z_1+z_2),
\end{eqnarray}
The inverse Green's function elements of Eq. (A1), can be rewritten so that the hermitian and anti-hermitian parts of self-energy, $\Sigma=\Sigma^L+\Sigma^R$, appear in inverse Green's function elements as follows
\begin{eqnarray}
   \nonumber G_{l,k}^{-1} &=& (E-\epsilon_{l})\delta_{l,k}-\frac{1}{2}\left(\Sigma_{l,k}(E)+\Sigma^{*}_{l,k}(E)\right) \\
  &-& \frac{1}{2}\left(\Sigma_{l,k}(E)-\Sigma^{*}_{l,k}(E)\right),
\end{eqnarray}
or
\begin{equation}\label{21}
   G_{l,k}^{-1}=\tilde{G}_{l,k}+\frac{i}{2}\Gamma_{l,k}
\end{equation}
in which we have used of
\begin{eqnarray}
  \Gamma_{l,k}(E)&=&i\left(\Sigma_{l,k}(E)-\Sigma^{*}_{l,k}(E)\right) , \\
  \tilde{G}_{l,k}&=&(E-\epsilon_{l})\delta_{l,k}-\frac{1}{2}\left(\Sigma_{l,k}(E)+\Sigma^{*}_{l,k}(E)\right)
\end{eqnarray}
and $\tilde{G}_{l,k}$ is a hermitian part and the last term in (B4) is the anti-hermitian part of $G^{-1}_{l,k}$. By using of Eqs. (A2)-(A7) and (A14), the above equation will be
\begin{eqnarray}
  \nonumber \Gamma_{l,k}(E)&=&\sum_{m,n,s}U_{l,m}\eta^{L,\dag}_{m,n}\eta^{L}_{n,s}U^{-1}_{s,k}\\
   &+&\sum_{m,n,s}U_{l,m}\eta^{R,\dag}_{m,n}\eta^{R}_{n,s}U^{-1}_{s,k}, \\
 \nonumber\tilde{G}_{l,k}&=&(E-\epsilon_{l})\delta_{l,k}-\Re g^{L}(E)\sum_{m}\tau_{m,l}^{L,*}\tau^{L}_{m,k}\\
 &-&\Re g^{R}(E)\sum_{m}\tau_{m,l}^{R,*}\tau^{R}_{m,k}
\end{eqnarray}

It is clear that both $\tilde{G}$ and $\Gamma$ are hermitian and so each trace is real and we have
\begin{eqnarray}
 \nonumber Tr(\tilde{G}) &=& 2E-(\epsilon_1+\epsilon_2)-2\Re g^L(E)\left(|\tau^{L}_{11}|^2+|\tau^{L}_{12}|^2\right)\\
 &-&2\Re g^R(E)\left(|\tau^{R}_{11}|^2+|\tau^{R}_{12}|^2\right)\\
  Tr(\Gamma)&=&|\eta^{L}_{11}|^2+|\eta^{L}_{12}|^2+|\eta^{R}_{11}|^2+|\eta^{R}_{12}|^2,
\end{eqnarray}
according to Eq.(B4) we have
\begin{equation}\label{21}
  Tr(G^{-1})=Tr(\tilde{G})+\frac{i}{2}Tr(\Gamma),
\end{equation}
and by using of Eqs.(B2) and (B9), after the comparison of both sides of above equation, one yields the following conditions
\begin{eqnarray}
\nonumber\Re(z_1+z_2)&=&(\epsilon_1+\epsilon_2)+2\Re g^L(E)\left(|\tau^{L}_{11}|^2+|\tau^{L}_{12}|^2\right) \hspace{.5cm}\\
  &+&2\Re g^R(E)\left(|\tau^{R}_{11}|^2+|\tau^{R}_{12}|^2\right) \\
\Im(z_1+z_2)&=& \frac{1}{2}\left(|\eta^{L}_{11}|^2+|\eta^{L}_{12}|^2+|\eta^{R}_{11}|^2+|\eta^{R}_{12}|^2\right)
\end{eqnarray}
If we consider the following definition for $z$ quantity
\begin{equation}\label{22}
z_{i}=\varepsilon_i+\frac{i}{2}(\gamma_i^L+\gamma_i^R)\ \ \ \ \ i=1,2
\end{equation}
where $\varepsilon_i$s are the energy of the two levels and $\gamma_i$s are the
broadening by contacts, then according to Eqs. (B12) and (B13) we reach
\begin{widetext}
\begin{eqnarray}
\varepsilon_1+\varepsilon_2&=& (\epsilon_1+\epsilon_2)+2\Re g^L(E)\left(|\tau^{L}_{11}|^2+|\tau^{L}_{12}|^2\right)
+2\Re g^R(E)\left(|\tau^{R}_{11}|^2+|\tau^{R}_{12}|^2\right)\\
  &&\frac{1}{2}\left(\gamma_1^L+\gamma_1^R+\gamma_2^L+\gamma_2^R\right)=  \frac{1}{2}\left(|\eta^{L}_{11}|^2+|\eta^{L}_{12}|^2+|\eta^{R}_{11}|^2+|\eta^{R}_{12}|^2\right)
   \end{eqnarray}
   \end{widetext}
   the above conditions indicate that the real part of self-energy causes shift in the system energy levels, while the imaginary part has the effect of level broadening.

   In the case of symmetric system-lead coupling, we take $\gamma_1^L=\gamma_1^R=\gamma_1/2$ and $\gamma_2^L=\gamma_2^R=\gamma_2/2$, and effective hybridization matrix elements should satisfy
   \begin{equation}\label{22}
     |\eta^{L}_{11}|^2=|\eta^{R}_{11}|^2,\ \ \ \ \ \ \ |\eta^{L}_{12}|^2=|\eta^{R}_{12}|^2
   \end{equation}
   so according to Eq. (B13) we obtain
   \begin{eqnarray}
     \gamma_1 &=& |\eta^{L}_{11}|^2=|\eta^{R}_{11}|^2, \\
     \gamma_2 &=& |\eta^{L}_{12}|^2=|\eta^{R}_{12}|^2
   \end{eqnarray}
   In what follows, we suppose that elements of effective hybridization matrix are real, so above relations can be written as
   \begin{eqnarray}
     \sqrt{\gamma_1}&=&\eta^{L}_{11}=\eta^{R}_{11}, \\
     \sqrt{\gamma_2}&=&\eta^{L}_{12}=\eta^{R}_{12}.
   \end{eqnarray}
   and equations (A7) and (A8) have explicit expression as
   \begin{equation}\label{23}
     T_i(E)=\frac{\gamma_i^2}{(E-\varepsilon_i)^2+\gamma_i^2},\ \ \ \ \ \ \ i=1, 2
   \end{equation}
   \begin{equation}\label{24}
     T_{12}(E)=2\gamma_1\gamma_2\frac{(E-\varepsilon_1)(E-\varepsilon_2)+\gamma_1\gamma_2}{\left((E-\varepsilon_1)^2+\gamma_1^2\right)\left((E-\varepsilon_2)^2+\gamma_2^2\right)}
   \end{equation}
   the above relation for $T_{12}$ can be expanded according to $T_{i}$s as
   \begin{equation}\label{25}
     T_{12}=\left[C_1ET_1(E)+C_2T_1(E)+D_1ET_2(E)+D_2T_2(E)\right]
   \end{equation}
   where constants $C_i$s have a following definitions
   \begin{eqnarray}
     C_1 &=& \frac{2\gamma_2(\varepsilon_1-\varepsilon_2)}{\gamma_1[(\varepsilon_1-\varepsilon_2))^2+(\gamma_1+\gamma_2)^2]} \\
     C_2 &=& 2\gamma_2\frac{\varepsilon_1\varepsilon_2+\gamma_1\gamma_2+\gamma_1^2-\varepsilon_1^2}{\gamma_1[(\varepsilon_1-\varepsilon_2))^2+(\gamma_1+\gamma_2)^2]}
       \end{eqnarray}
   and constants $D_i$s will obtain by applying  $\varepsilon_1\leftrightarrow\varepsilon_2$ and also $\gamma_1\leftrightarrow\gamma_2$ exchanges in corresponding $C_i$s.
   Finally, we can express transmission function $T(E)$ as follows
   \begin{equation}\label{26}
     T(E)=(1+ C_1E+ C_2)T_1(E)+(1+ D_1E+ D_2)T_2(E),
\end{equation}
\section{Heat dissipation}
According to the Landauer approach, the electronic contribution to the charge current ($I$) and the energy current ($I_E$) through the junction can be expressed in terms of the transmission function for positive bias $\mu_{L}-\mu_{R}=eV$, as
\begin{eqnarray}
 I(V) &=&\frac{2e}{h}\int_{-\infty}^{+\infty}T(E,V)F(E,\mu_L,\mu_R)dE, \\
 I_E(V) &=&\frac{2e}{h}\int_{-\infty}^{+\infty}ET(E,V)F(E,\mu_L,\mu_R)dE
 \end{eqnarray}
 where $F(E,\mu_L,\mu_R)=f_L(E,\mu_L)-f_R(E,\mu_R)$ and $f_{L(R)}(E,V)$ is the Fermi function of the left (right) electrode. Each Fermi function depends on the electrode's chemical potential, which in turn is related to the applied bias. The rate of heat released  in the left (right) electrode with electrochemical potential $\mu_{L(R)}$ is given by
 \begin{equation}\label{34}
  Q_{L(R)}=\frac{\mu_{L(R)}}{e}I-I_E,
 \end{equation}
 Using Eq. (C1) and (C2) we obtain the rate of heat (heat per unit of time) dissipated in left and right electrodes as
\begin{eqnarray}
  \nonumber Q_{L}&=&\frac{2}{h}\int_{-\infty}^{+\infty}(\mu_{L}-E)T(E,V)F(E,\mu_L,\mu_R)dE \\
   Q_{R}&=&\frac{2}{h}\int_{-\infty}^{+\infty}(E-\mu_{R})T(E,V)F(E,\mu_L,\mu_R)dE
\end{eqnarray}
thus total heat dissipation (heat per unit of time) in the system is
\begin{eqnarray}
\nonumber Q_{Total}(V)&=&Q_{L}(V)+Q_{R}(V) \\
\nonumber &=& \frac{2eV}{h}\int_{-\infty}^{+\infty}T(E,V)F(E,\mu_L,\mu_R)dE=IV \\
\end{eqnarray}
Three limits allow simplifications of the above relation. First, is the limit that two electrodes to be of the same materials, $\mu_L=\mu_R=\mu$, and the system is in equilibrium at zero bias and without loss of generality we set $\mu=0$. Moreover, we assume that the electrochemical potentials are shifted symmetrically with the bias
voltage, i.e. $\mu_L = eV/2$ and $\mu_R = -eV/2$. Second, is the limit of zero temperature, $f_{L(R)}(E,V)\rightarrow \Theta(-E\pm eV)$, where $\Theta(x)$ is
the Heaviside (step) function. The use of plus (minus) signs here
arises from the positive (negative) bias. Third, is the low bias limit, where we can neglect
the transmission function's dependence on the bias, $T(E,V)\rightarrow T(E)$. Applying these limits, we arrive
\begin{eqnarray}
 Q_L(V)&=&\frac{2}{h}\int_{-eV/2}^{+eV/2}(eV/2-E)T(E)dE,\\
 Q_R(V)&=&\frac{2}{h}\int_{-eV/2}^{+eV/2}(E+eV/2)T(E)dE, \\
 Q_{Total}(V)&=&\frac{2eV}{h}\int_{-eV/2}^{+eV/2}T(E)dE,
\end{eqnarray}
If one concentrates on the bias polarity, it is easy to show that
      \begin{equation}\label{23}
   Q_{L(R)}(V)=Q_{R(L)}(-V),
 \end{equation}
 The above relation indicates that the power dissipated in one of the electrodes can be obtained from
the power dissipated in the other one by simply inverting the bias. If one checks the behaviour of total
 heat dissipation with respect to the inversion of the bias voltage, one obtains
\begin{eqnarray}
  \nonumber Q_{Total}(-V) &=& Q_L(-V)+Q_R(-V) \\
   &=&Q_{R}(V)+Q_{L}(V)=Q_{Total}(V)
\end{eqnarray}
thus, the total heat dissipation is a symmetric function with respect to the inversion of the bias voltage.
Also, for total heat we have
\begin{eqnarray}
\nonumber Q_{Total}(V) &=& Q_{L(R)}(V)+Q_{R(L)}(V) \\
&=& Q_{L(R)}(V)+Q_{L(R)}(-V)
\end{eqnarray}
this relation implies
that the total power dissipated in the model is equal to the sum of the power dissipated for positive and
negative bias in a given electrode.


\vspace{0.3cm}
\section*{References}
\begin{thebibliography}{10}
\bibitem{ree} M.A. Reed, C. Zhou, C. J. Muller, T.P. Burgin, and J. M. Tour, Science \textbf{278}, 252 (1997).
\bibitem{red} P. Reddy, S.Y. Jang, R. A. Segalman, and A. Majumdar, Science \textbf{315}, 1568 (2007).
\bibitem{mai} S. K. Maiti, J. Nanosci.Nanotechnol. \textbf{8} 4096 (2008).
\bibitem{saf} A. Saffarzadeh, J. Appl.Phys. \textbf{103} 083705 (2008).
\bibitem{ker} C. Kergueris, J.P. Bourgoin, D. Esteve, C. Urbina, M. Magoga,
and C. Joachim, Phys. Rev. B \textbf{59}, 12505 (1999).
\bibitem{por} D. Porath, A. Bezryadin, S. de. Vries, and C. Dekker, Nature
\textbf{403}, 635 (2000).
\bibitem{rei} J. Reichert, R. Ochs, D. Beckmann, H. B. Weber, M.
Mayor, and H. v. Lohneysen, Phys. Rev. Lett. \textbf{88}, 176804
(2002).
\bibitem{nit} A. Nitzan and M. A. Ratner, Science \textbf{300}, 1384 (2003).
\bibitem{pau} M. Paulsson and S. Datt, Phys. Rev. B \textbf{67} 241403 (2003)
\bibitem{paul} F. Pauly, J. K. Viljas and J. C. Cuevas, Phys. Rev. B \textbf{78} 035315 (2008)
\bibitem{ke} S. H. Ke, W. Yang, S. Curtarolo and H. U. Baranger, Nano Lett. \textbf{9} 1011 (2009)
\bibitem{fin} C. M. Finch, V. M. Garc\'{\i}a-Su\'{a}rez and C.J. Lambert, Phys. Rev. B \textbf{79} 033405 (2009)
\bibitem{liu} Y. S. Liu and Y. C. Chen, Phys. Rev. B \textbf{79} 193101 (2009)
\bibitem{berg1} J. P. Bergfield and C. A. Stafford, Nano Lett. \textbf{9} 3072 (2009)
\bibitem{berg2} J. P. Bergfield, M. A. Solis and C. A. Stafford, ACS Nano \textbf{4} 5314 (2010)
\bibitem{noz} D. Nozaki, H. Sevinçlic, W. Li, R. Guti\'{e}rrez and G. Cuniberti, Phys. Rev. B \textbf{81} 235406 (2010)
\bibitem{que} S. Y. Quek, H. J. Choi, S. G. Louie and J. B. Neaton, ACS Nano \textbf{5} 551 (2011)
\bibitem{ser} N. Sergueev , S. Shin, M. Kaviany and B. Dunietz, Phys. Rev. B \textbf{83} 195415 (2011)
\bibitem{stad} R. Stadler and T. Markussen, J. Chem. Phys. \textbf{135} 154109 (2011)
\bibitem{saha} K. K. Saha, T. Markussen, K. S. Thygesen and B. K. Nikolic, Phys. Rev. B \textbf{84} 041412 (2011)
\bibitem{bil} S. Bilan, L. A. Zotti, F. Pauly and J. C. Cuevas, Phys. Rev. B \textbf{85 } 205403 (2012)
\bibitem{bur} M. B\"{u}rkle, L. A. Zotti, J. K. Viljas, D. Vonlanthen, A. Mishchenko, T. Wandlowski, M. Mayor M, G. Sch\"{o}n and
F. Pauly , Phys. Rev. B \textbf{86} 115304 (2012)
\bibitem{mar} T. Markussen, C. Jin and K. S. Thygesen, Phys. Status Solidi b \textbf{250} 2394 (2013)
\bibitem{bah} K. Baheti, J. A. Malen, P. Doak, P. Reddy, S. Y. Jang, T. D. Tilley, A. Majumdar and R. A. Segalman, Nano Lett. \textbf{8} 715 (2008)
\bibitem{mal} J. A. Malen, P. Doak, K. Baheti, T. D. Tilley, R. A. Segalman and A. Majumdar, Nano Lett.\textbf{ 9} 1164 (2009)
\bibitem{tan} A. Tan, S. Sadat and P. Reddy, Appl. Phys. Lett. \textbf{96} 013110 (2010)
\bibitem{yee} S. K. Yee, J. A. Malen, A. Majumdar and R. A. Segalman, Nano Lett. \textbf{11} 4089 (2011)
\bibitem{tan2} A. Tan, J. Balachandran, S. Sadat, V. Gavini, B. D. Dunietz, S. Y. Jang and P. Reddy, J. Am. Chem. Soc.
\textbf{133} 8838 (2011)
\bibitem{wid} j. R. Widawsky, P. Darancet, J. B. Neaton and L. Venkataraman, Nano Lett. \textbf{12} 354 (2011)
\bibitem{eva} C. Evangeli, K. Gillemot, E. Leary, M. T. González, G. Rubio-Bollinger, C. J. Lambert and N. Agra\"{\i}t, Nano
Lett. \textbf{13} 2141 (2013)
\bibitem{wid2} J. R. Widawsky, W. Chen, H. V\'{a}zquez, T. Kim, R. Breslow, M. S. Hybertsen and L. Venkataraman,  Nano Lett.
\textbf{13} 2889 (2013)
\bibitem{lee} W. Lee, K. Kim, W. Jeong,L. A. Zotti, F. Pauly F, J. C.
Cuevas and P. Reddy, Nature \textbf{489}, 209 (2013)
\bibitem{zot} L. A. Zotti, M. Bürkle, F. Pauly, W. Lee, K. Kim, W. Jeong, Y. Asai, P. Reddy and J. C. Cuevas, New Journal of Physics {\bf 16}, 015004 (2014)
    \bibitem{butt} M. B\"{u}ttiker, Y. Imry, R. Landauer, and S. Pinhas, Phys. Rev. B {\bf31}, 6207 (1985).
    \bibitem{kim} Y. Kim, W. Jeong,	K. Kim,	W. Lee, P. Reddy, Nature Nanotechnology {\bf9}, 881 (2014)
    \bibitem{lu} X. Lu, M. Grobis, K. H. Khoo, Steven G. Louie, and M. F. Crommie, Phys. Rev. B {\bf70}, 115418 (2004)
    \bibitem{dif} We set two energy levels of toy model from HOMO and LUMO of C$_{60}$ because we want to compare the behaviour of toy model and C$_{60}$ molecular junctions.
    \bibitem{gera} G. G\'{e}ranton, C. Seiler, A. Bagrets, L. Venkataraman and F. Evers, J. Chem. Phys. {\bf139}, 234701 (2013)
        \bibitem{noz2} D. Nozaki, H. M. Pastawski, and G. Cuniberti, New J. Phys. {\bf12}, 063004 (2010).
\end{thebibliography}
\end{document}